\documentclass[conference]{IEEEtran}
\usepackage{subfigure}
\usepackage{graphicx}
\usepackage{amsmath}
\usepackage{bm}
\usepackage{url}
\usepackage{stmaryrd}
\usepackage{balance}
\usepackage{amssymb}
\usepackage{pgfplots}
\usepackage{pifont}
\usepackage[usenames,dvipsnames]{pstricks}
\usepackage{epsfig}
\usepackage{booktabs}
\usepackage{pgffor}
\usepackage{tikz}
\usepackage{multirow}
\ifCLASSOPTIONcompsoc
  \usepackage[nocompress]{cite}
\else
  \usepackage{cite}
\fi

\newcommand{\our}[0]{\textsc{Scan-ps}}
\newcommand{\ourb}[0]{\textsc{Scan}}
\newcommand{\ourd}[0]{\textsc{Source}}
\newcommand{\ourf}[0]{\textsc{Old-ps}}
\newcommand{\ourh}[0]{\textsc{Old}}
\newcommand{\ouri}[0]{\textsc{New}}
\newcommand{\ournaive}[0]{\textsc{Naive}}
\newcommand{\ourtrad}[0]{\textsc{Trad}}
\newcommand{\ourtradsp}[0]{\textsc{Trad-ps}}

\begin{document}
\title{Predicting Social Links for New Users across Aligned Heterogeneous Social Networks}

\author{

\IEEEauthorblockN{Jiawei Zhang}
\IEEEauthorblockA{University of Illinois at Chicago\\
Chicago, IL, USA\\
jzhan9@uic.edu}

\and

\IEEEauthorblockN{Xiangnan Kong}
\IEEEauthorblockA{University of Illinois at Chicago\\
Chicago, IL, USA\\
xkong4@uic.edu}

\and

\IEEEauthorblockN{Philip S. Yu}
\IEEEauthorblockA{University of Illinois at Chicago\\
Chicago, IL, USA\\
psyu@cs.uic.edu}
}

\maketitle

\begin{abstract}
Online social networks have gained great success in recent years and many of them involve multiple kinds of nodes and complex relationships. Among these relationships, social links among users are of great importance. Many existing link prediction methods focus on predicting social links that will appear in the future among all users based upon a snapshot of the social network. In real-world social networks, many new users are joining in the service every day. Predicting links for new users are more important. Different from conventional link prediction problems, link prediction for new users are more challenging due to the following reasons: (1) differences in information distributions between new users and the existing active users (i.e., old users); (2) lack of information from the new users in the network. In order to solve the above problems, we need to accommodate the differences in information distributions between old users and new users within the network and transfer additional information from other sources for the new users. We notice that users nowadays are normally involved in multiple social networks to enjoy more online services at the same time, such as Facebook, Twitter and Foursquare. New users in one social network (e.g., Foursquares) might have already joined and been active in another network (e.g., Twitter) for a long time. We propose a link prediction method called {\our} (Supervised Cross Aligned Networks link prediction with Personalized Sampling), to solve the link prediction problem for new users with information transferred from both the existing active users in the target network and other source networks through aligned accounts. We proposed a within-target-network personalized sampling method to process the existing active users' information in order to accommodate the differences in information distributions before the intra-network knowledge transfer. {\our} can also exploit information in other source networks, where the user accounts are aligned with the target network. In this way, {\our} could solve the cold start problem when information of these new users is total absent in the target network. Extensive experiments conducted on Twitter and Foursquare, two real-world aligned heterogeneous social networks, demonstrate that {\our} outperforms other link prediction methods for new users under different degrees of newness consistently and works well with the cold start problems.
\end{abstract}

\section{Introduction}
\label{sec:intro}

Online social networks are becoming more and more popular in recent years. Many of these networks involve multiple kinds of nodes, such as  users, posts, locations, et al.,  and complex relationships among the nodes, such as social links and location check-ins. Among these relationships, social link prediction is crucial for many social networks because it will lead to more connections among users. Meanwhile, well-established online social relationships will attract users to use the network more frequently \cite{KLPM10}.


Many of previous works on link prediction focus on predicting potential links that will appear among all the users, based upon a snapshot of the social network. These works treat all users equally and try to predict social links for all users in the network. However, in real-world social networks, many new users are joining in the service every day. Predicting social links for new users are more important than for those existing active users in the network as it will leave the first impression on the new users. First impression often has lasting impact on a new user and may decide whether he will become an active user. A bad first impression can turn a new user away. So it is important to make meaningful recommendation to  a new user to create a good first impression and attract him to participate more. For simplicity, we refer users that have been actively using the the network for a long time as ``old users''. It has been shown in previous works that there is a negative correlation between the age of nodes in the network and their link attachment rates. The distribution of linkage formation probability follows a power-law decay with the age of nodes \cite{KE02}. So, new users are more likely to accept the recommended links compared with existing old users and predicting links for new users could lead to more social connections. 

\begin{figure}[!t]
 \centering    
 \begin{minipage}[c]{0.8\columnwidth}
  \centering
    \includegraphics[width=1.0\textwidth]{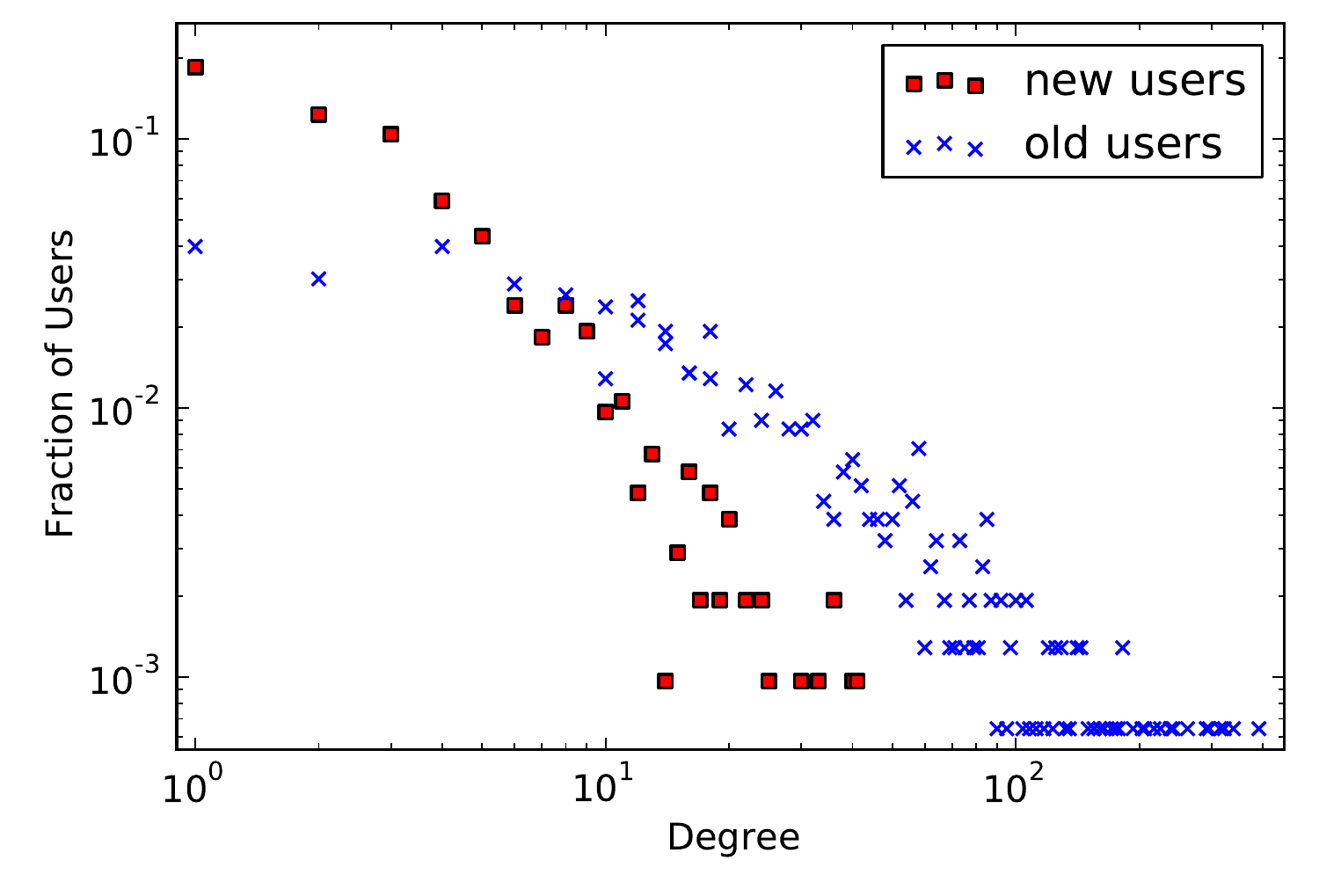}
 \end{minipage}
\caption{Degree distributions of users in Foursquare network.}\label{fig_eg_f}
\end{figure}

\begin{figure}[!t]
 \centering    
 \begin{minipage}[c]{0.8\columnwidth}
  \centering
    \includegraphics[width=1.0\textwidth]{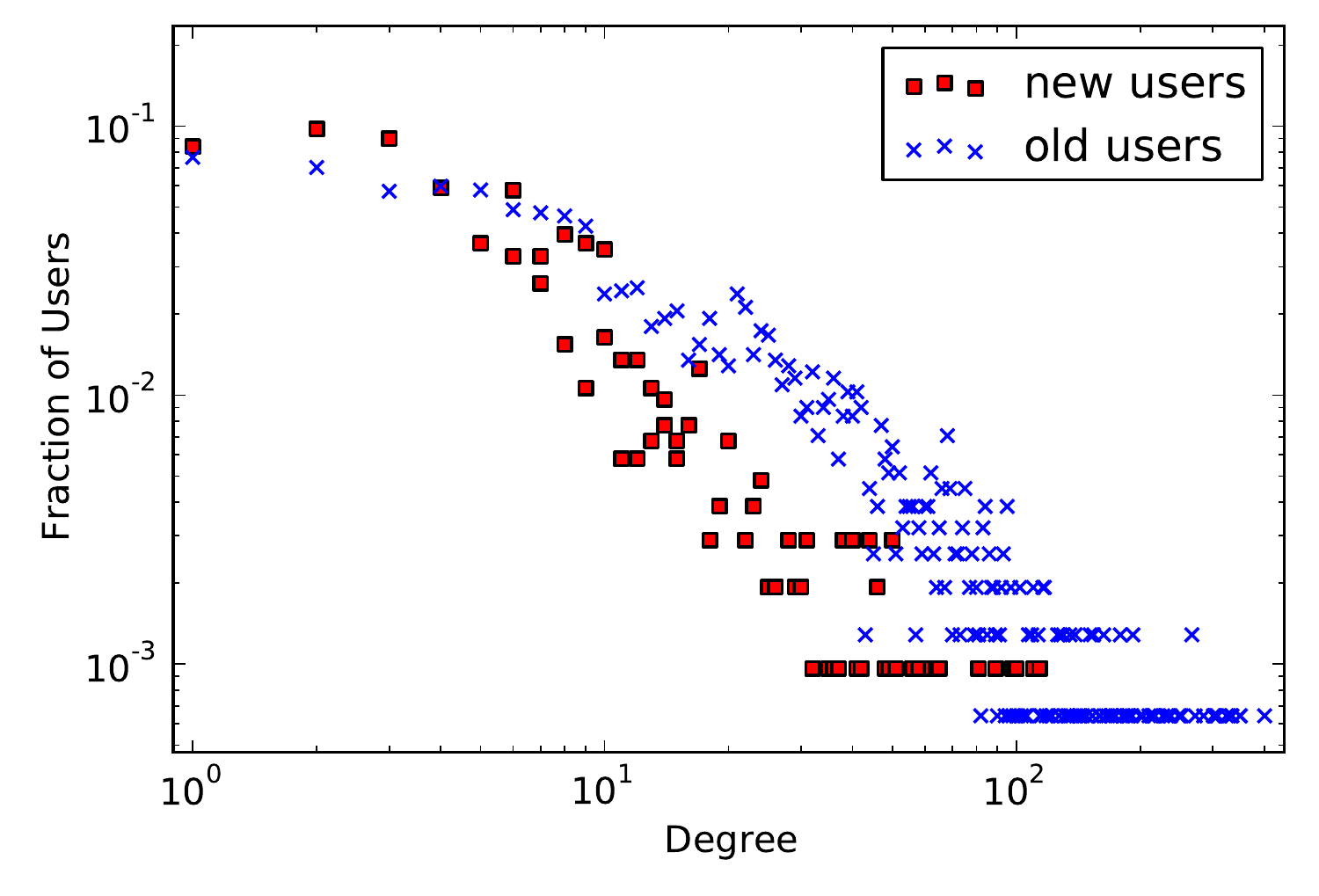}
 \end{minipage}
\caption{Degree distributions of users in Twitter network.}\label{fig_eg_t}
\end{figure}

In this paper, we study the problem of predicting social links for new users, who have created their accounts for just a short period of time. The link prediction problem for new users is different from traditional link prediction problems. Conventional supervised link prediction methods implicitly or explicitly assume that the information are identically distributed over all the nodes in the network without considering the joining time of  the users. The models trained over one part of the network can be directly used to predict links in other parts of the network. However, in real-world social networks, the information distributions of the new users could be very different from old users. New users may have only a few activities or even no activities (i.e., no social links or other auxiliary information) in the network. While, old users usually have abundant activities and auxiliary information in the network. In Figures~\ref{fig_eg_f} and ~\ref{fig_eg_t}, we show the degree distributions of users who registered their accounts within three months and old users who registered more than three months before in Twitter and Foursquare respectively. We find that the social link distributions of new users and old users are totally different from each other in both Foursquare and Twitter. As a result, conventional supervised link prediction models trained over old users based upon structural features, such as common neighbors, may not work well on the new  users.

\begin{figure}[!t]
 \centering    
 \begin{minipage}[l]{0.8\columnwidth}
  \centering
    \includegraphics[width=1.0\textwidth]{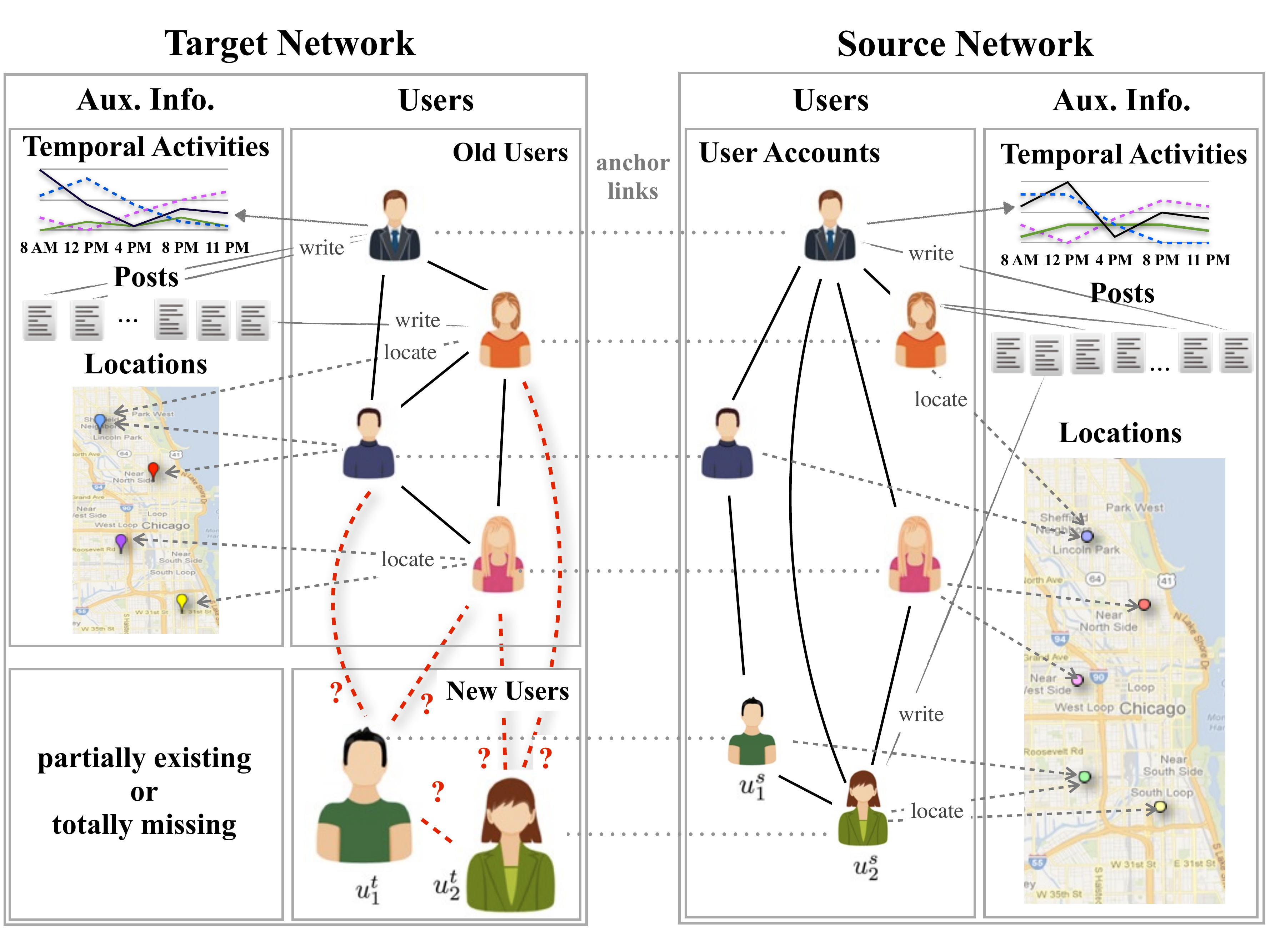}
 \end{minipage}
\caption{Example of predicting social links across two aligned heterogeneous online social networks.}\label{fig_eg1}
\end{figure}

Another challenging problem in link prediction for new users is that information owned by new users can be very rare or even totally missing. Conventional methods based upon one single network will not work well due to the lack of historical data about the new users. In order to solve this problem, we need to transfer additional information about the new users from other sources. Nowadays, people are usually involved in multiple social networks to enjoy more services. For example, people will join Foursquare to search for nearby restaurants to have dinner with their family. Meanwhile, they tend to use Facebook to socialize with their friends and involve in Twitter to post comments about recent news. The accounts of the same user in different networks can be linked through account alignments. For example, when users register their Foursquare accounts, they can use their Facebook or Twitter accounts to sign in the Foursquare  network. In this paper, we name such links among accounts of the same user as ``anchor links" \cite{KZY13}, which could help align user' accounts across multiple social networks. For example, in Figure~\ref{fig_eg1}, there are many users in two networks respectively. We find that the accounts in these two networks are actually owned by $6$ different users in reality and we add an \textit{anchor link} between each pair of user accounts corresponding to the same user. Via the \textit{anchor links}, we could locate users' corresponding accounts in the other networks. 

New users in one social network (i.e., \textit{target} network) might have been using other social networks (i.e., \textit{source} networks) for a long time. These user accounts in the source networks can provide additional information about the new users in the source network. This additional information is crucial for link prediction for new users, especially when the new users have little activities or no activities in the target network (i.e., cold start problem). For example, in Figure~\ref{fig_eg1}, we have two social networks, i.e., the target network and the source network, with aligned user accounts. In the target network, there are many old users with abundant social links and auxiliary information, such as posts, spatial and temporal activities. In addition, there are also some new users, i.e., user $u^t_1$ and $u^t_2$, in the target network. These two new users have just created their accounts in the target network and have not yet created many social links or auxiliary information. However, we can see that there is abundant information about these two new users in the source network, because of their ``anchor linked" user accounts $u_1^s$ and $u_2^s$ in the source network. In this paper, we propose to exploit the new users' information in source networks to help improve the link prediction performances in the target network.



\begin{figure*}[t]
\centering
\subfigure[Link Prediction via Biased Sampling]{ \label{fig_eg2_1}
    \begin{minipage}[l]{.60\columnwidth}
       \centering
      \includegraphics[width=1.0\textwidth]{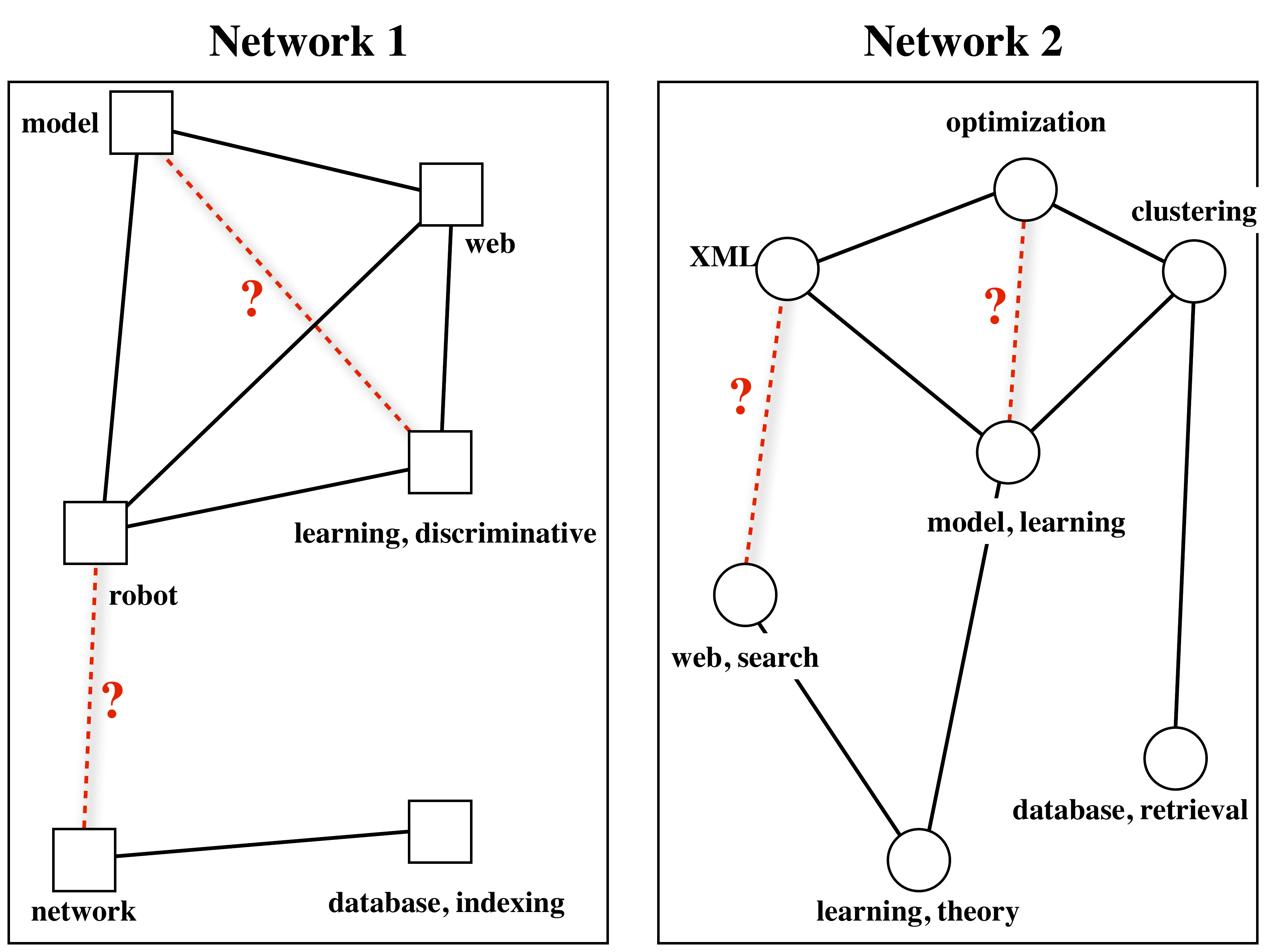}
    \end{minipage}
}
\subfigure[Pseudo Cold Start Link Prediction]{ \label{fig_eg2_2}
    \begin{minipage}[l]{.60\columnwidth}
      \centering
      \includegraphics[width=1.0\textwidth]{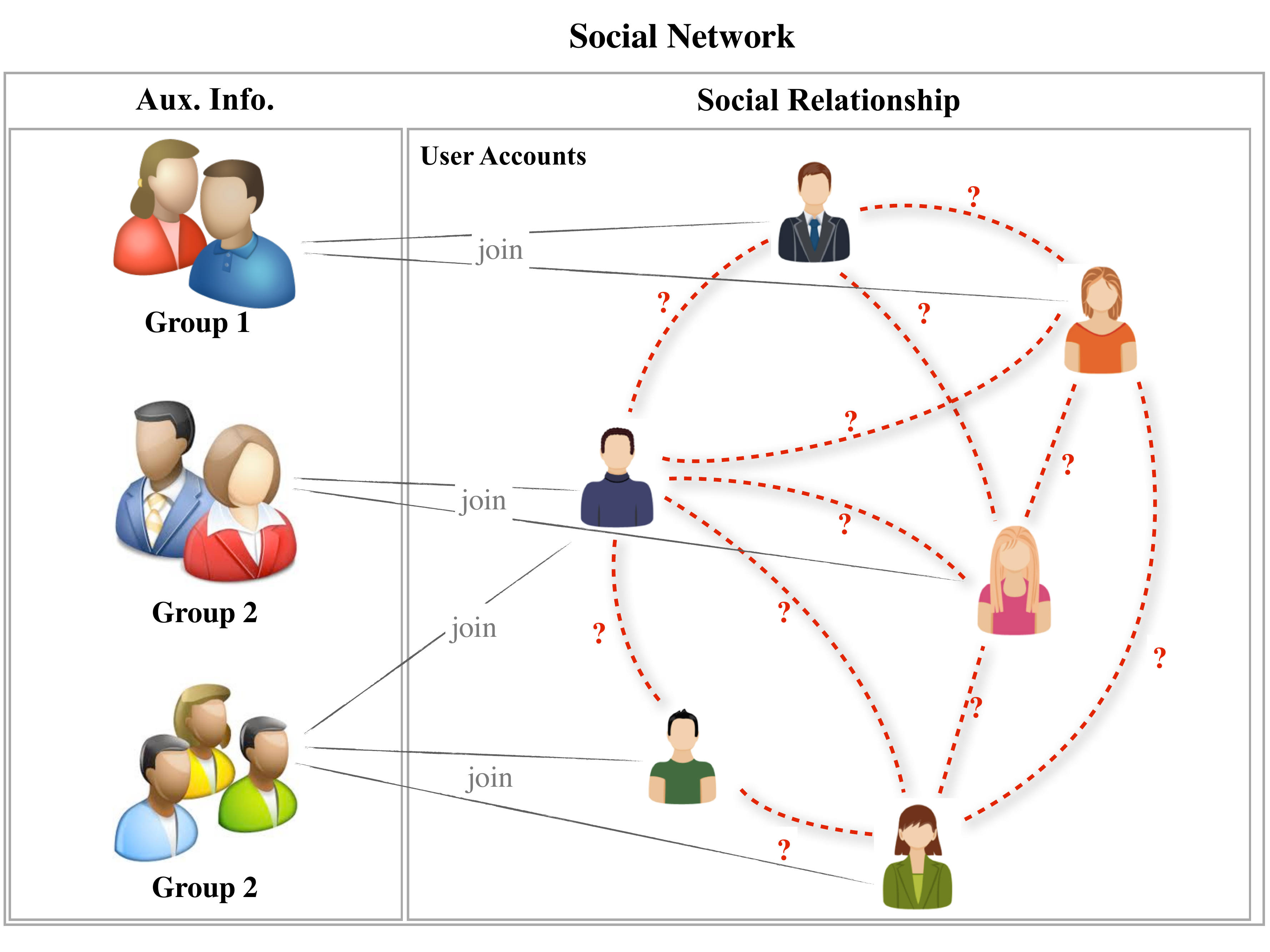}
    \end{minipage}
}
\subfigure[Link Prediction via Network Transfer]{ \label{fig_eg2_3}
    \begin{minipage}[l]{.60\columnwidth}
      \centering
      \includegraphics[width=1.0\textwidth]{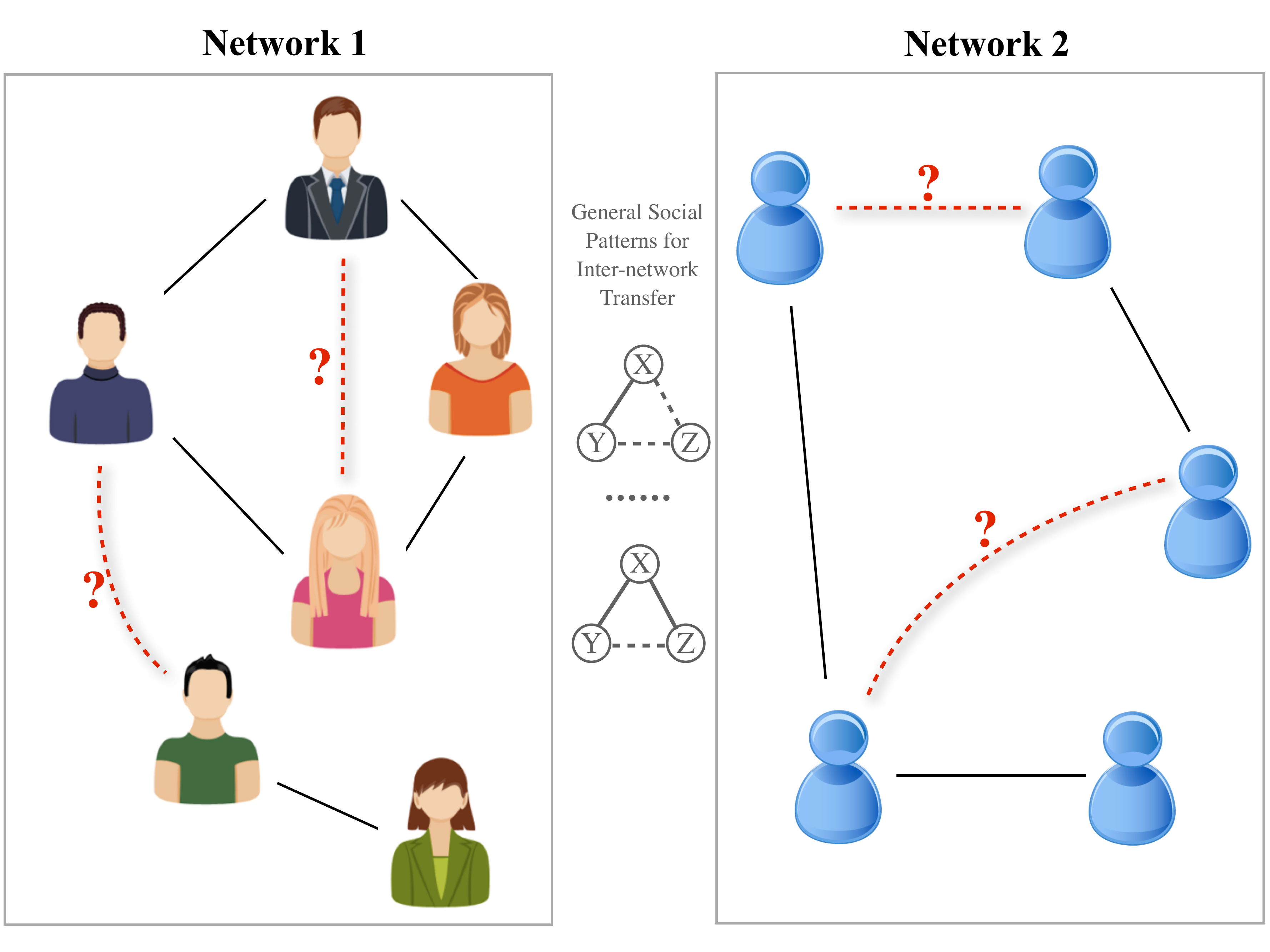}
    \end{minipage}
}
\caption{Comparison of three other different link prediction problems}\label{fig_eg2}
\end{figure*}

The problem of social link prediction for new users by using aligned social networks has not been studied yet. It is novel and totally different from existing link prediction problems, e.g., link prediction via personalized sampling \cite{QAH13}, pseudo cold start link prediction problem \cite{GZ12, LCB10, ZLZZ09}, link prediction via network transfer \cite{LSTD10, TLK12} and the traditional transfer learning problems in feature space \cite{CTT11, DCXYY09, LYX09, LZXXH08, XDXY07}. A more detailed comparison of all these problems are shown in Table~\ref{tab:related} and Figure~\ref{fig_eg2}.

\begin{table*}[t]
\centering
\caption{Summary of related problems.}\label{tab:related}
\begin{tabular}{|l|c|c|c|c|c|}
\hline
& \textbf{Predicting Links}
& \textbf{Link Prediction}
& \textbf{Pseudo Cold Start}
& \textbf{Link Prediction}
& \textbf{Transfer Learning}\\
& \textbf{across Aligned Networks}
& \textbf{via Biased}
& \textbf{Link Prediction}
& \textbf{via Network}
& \textbf{in Feature Space}\\
Property
& \textbf{with Personalized Sampling}
& \textbf{Sampling \cite{QAH13}}
& \textbf{\cite{GZ12, LCB10, ZLZZ09}}
& \textbf{Transfer\cite{LSTD10, TLK12}}
& \textbf{\cite{CTT11, DCXYY09, LYX09, LZXXH08, XDXY07}}\\
\hline
\# networks	&multiple		&multiple	&single			&multiple	&multiple domains \\
\hline
network type	&heterogeneous	&homogeneous	&heterogeneous	&heterogeneous	&heterogeneous \\
\hline
network alignment	& yes	&no	&no	&no	&n/a \\
\hline
target	&new users	&incomplete network	&incomplete network	&incomplete network	&n/a \\
\hline
sampling	&within network	&across network	&n/a	&n/a	&n/a \\
\hline
can handle cold &cold start	& n/a	&pseudo cold start	&n/a	&n/a \\
start problem? & & & & &\\
\hline
knowledge		& network structure	&network structure	&n/a		&triad patterns	 in &knowledge in \\
to transfer &through anchor links	&via attribute info. &&network structure&feature space \\
\hline
transfer route &intra-network &inter-network&n/a &inter-network &intra-domain \\
&\textbf{and} inter-network&&&&\textbf{or} inter-domain\\
\hline
\end{tabular}
\end{table*}

In spite of its significance, social link prediction for new users across aligned social networks is very challenging to solve due to the following reasons:
\begin{enumerate}
 \item \textit{Differences in information distributions}. In order to use the old users' information in the target network, we need to overcome the problem of the differences in information distributions between old users and new users.
 \item \textit{No auxiliary information}. Another key part of the problem we want to study is the cold start link prediction problem caused by the lack of information about these new users. We need to find other information sources for such problem and provide the new users with high-quality social link prediction results.
 \item \textit{Aligned social networks}. Previous works on transfer learning focus on transferring knowledge between two domains via shared feature space or between two networks through shared triad linkage structures. No works have been done on aligned social networks yet.
\end{enumerate}

In order to solve these problems, we propose a novel supervised cross aligned networks link recommendation method, {\our}. Different from previous works, {\our} extracts heterogeneous features from other aligned networks to improve link prediction results for new users in the target network. We analyze the problem about the differences in information distributions between new users and old users in details and propose a within-network personalized sampling method to accommodate that difference. What's more, {\our} could also solve the cold start social link prediction problem assisted by other aligned source networks. Intra and inter network information transfers are conducted simultaneously to make full use of the information contained in these aligned networks to improve the prediction result.

\section{Problem Formulation}
\label{sec:formulation}
The problem studied in this paper is social link prediction for new users. We propose a supervised method based on aligned heterogeneous networks. In this section, we first define the concept of heterogeneous social network and aligned heterogeneous networks and then present the formulation of the social link prediction problem.

\noindent\textbf{Definition 1} (Heterogeneous Networks): Let $G = (V, E)$ be a network containing different kinds of information, where the set $V = \bigcup_i{V_i}$ contains multiple kinds of nodes, where $V_i, i \in \{1, 2, \cdots, |V|\}$ is the set of nodes of the same kind, $E = \bigcup_i{E_i}$ contains multiple types of links among the nodes, where $E_i, i \in \{1, 2, \cdots, |E|\}$ is the set of links of the same type.

In this paper, a heterogeneous social network is defined as $G = (V, E)$, where $V = U \cup L \cup T \cup W$ is the set of nodes in the network and $E = E_{social} \cup E_{locate} \cup E_{time} \cup E_{word}$ is the set of heterogeneous links in the network. $U, L, T$ and $W$ are the set of users, locations, time and word respectively, while $E_{social}, E_{locate}, E_{time}$ and $E_{word}$ are the sets of links between friends, locations, timestamps, words and users in $U$

\noindent\textbf{Definition 2} (Aligned Heterogeneous Networks): Let $\mathcal{G} = (G_{set}, A_{set})$ be aligned heterogeneous social networks, where $G_{set} = \{G^1, G^2, \cdots, G^n\}$ is the set of heterogeneous social networks, whose size is $n = \left|G_{set}\right|$, and $A_{set} = \{A^{1,2}, A^{1,3}, \cdots, A^{1,n}, A^{2,1}, \cdots, A^{n,n-1}\}$ is the set of directed anchor links between pairwise networks in $G_{set}$ and $A^{i,j} \subseteq U^i \times U^j$ is the set of anchor links between $G^i$ and $G^j$, where $U^i$ and $U^j$ are the user sets in graph $G^i$ and $G^j$.

\noindent \textbf{Definition 3}(Anchor Link): Link $(u_m^i, u_n^j)$ is an \textit{anchor link} between $G^i$ and $G^j$ iff. ($u_m^i \in U^i) \land (u_n^j \in U^j) \land (u_m^i$ and $u_n^j$ are accounts owned by the same user).


\textbf{Social Link Recommendation} is a traditional problem first proposed by \cite{LK03} and has been studied for many years. Different from prior works, in this paper, we want to study this problem for new users in the target network by using aligned heterogeneous social networks. Let $\mathcal{G} = (\{G^t, G^s\}, \{A^{t,s}, A^{s,t}\})$ be two aligned heterogeneous social networks, where $G^t$ is the target network and $G^s$ is an aligned source network. $A^{t,s}$ and $A^{s,t}$ are the sets of directed anchor links between $G^t$ and $G^s$. We want to predict social links for the new users in the target network. Let $U^t = U_{new}^t \cup U_{old}^t$  be the user set in $G^t$, where $U_{new}^t, U_{old}^t$ are the sets of new users and old users and $U_{new}^t \cap U_{old}^t = \emptyset$. What we want to predict is a subset of potential social links between the new users and all other users: $L \subseteq U_{new}^t \times U^t$. In other words, we want to build a function $f: L \rightarrow \{0, 1\}$, which could decide whether certain links related to new users exist in the target network or not.

\section{Proposed Methods}
In this section, we will describe the sets of features extracted from the aligned heterogeneous networks first. Next, we will analyze the problem of the differences in information distributions between old users and new users in the target network and propose a personalized sampling method to solve it. Finally, we will propose our supervised cross network link prediction method {\our} to predict potential social links for new users in the target network with information in aligned heterogeneous networks.

\subsection{Heterogeneous Features}
The networks used by us are heterogeneous and involve different nodes and complex dependency relationships. To make full use of information inside the networks, we extract four different categories of features, which are (1) social features, (2) spatial distribution features, (3) temporal distribution features and (4) text content features. More detailed information about these features is listed below.

\begin{itemize}
 \item \textbf{Social Features}: Social relationships are what we intend to predict for new users and the existing social information in the network should be utilized definitely. We extract three different social features from the social information in the networks, which are `` \textit{common neighbour} " (CN), ``\textit{Jaccard's Coefficient}" (JC) and ``\textit{Adamic/Adar Measure}" (AA) \cite{AA01}.\\
\textbf{Common neighbour}: $CN(u_i, u_j)$ shows the number of shared neighbours of user $u_i$ and $u_j$ in the network. Let $\Gamma(u)$ denote the set of neighbours of user $u$ in the network, then:
$$CN(u_i, u_j) = \left| \Gamma(u_i) \cap \Gamma(u_j) \right|$$
\textbf{Jaccard's coefficient}: $JC(u_i, u_j)$ takes all the neighbours of $u_i$ and $u_j$ into account, considering that $CN(u_i, u_j)$ could be very large because each one has a lot of neighbours rather than they are strongly related to each other. 
$$JC(u_i, u_j) = \frac{\left| \Gamma(u_i) \cap \Gamma(u_j) \right|}{\left| \Gamma(u_i) \cup \Gamma(u_j) \right|}$$
\textbf{Adamic/Adar Measure}: $AA(u_i, u_j)$ further gives each common neighbour $u_k$ a weight  $\frac{1}{\log{\left| \Gamma(u_k)\right|}}$ to measure $u_k$'s significance, which is defined as follows:
$$AA(u_i, u_j) = \sum_{u_k\in(\Gamma(u_i)\cap\Gamma(u_j))}{\frac{1}{\log{\left| \Gamma(u_k)\right|}}}$$

 \item \textbf{Spatial Distribution Features}: Lots of social networks have provided users the service to allow users to display their locations along with the online posts. We organize the locations that two users $u_i$ and $u_j$ have visited in two location vectors: \textit{l}$(u_i)$ and \textit{l}$(u_j)$. These two vectors are of the same length and each column corresponds to a specific location. The values stored in these two vectors are the times that the users has visited specific places. We extract several features from the location information: (1) inner product of these two location vectors, which is $l(u_i) \cdot l(u_j)$, (2) cosine similarity of these two location vectors: $\frac{l(u_i) \cdot l(u_j)}{{\left\|l(u_i)\right\|}\cdot{\left\|l(u_j)\right\|}}$, which takes the total number of locations that these two users have been to into account, (3) the Euclidean distance of two users' location vectors: $(\sum_k{(l(u_i)_k - l(u_j)_k)^2})^{1/2}$. In addition,  we also extend CN and JC to get the number and ratio of shared locations that two users have both visited, which are (4) $\textsc{CNlocation}(u_i, u_j) = \left| \Phi(u_i) \cap \Phi(u_j) \right|$ and (5) $\textsc{JClocation}(u_i, u_j) = \frac{\left| \Phi(u_i) \cap \Phi(u_j) \right|}{\left| \Phi(u_i) \cup \Phi(u_j) \right|}$, where $\Phi(u)$ denotes the set of locations that user $u$ has visited. (6) The average geographic distance of all the locations that users $u_i$ and $u_j$ have been to is also used as a feature.
 \item \textbf{Temporal Distribution Features}: Besides the location information, we could also get the timestamps of online posts, which could reveal users' activity patterns. We divide each day into 24 slots and save the number of posts that a user posted at each hour in vector $t(u)$, whose length is 24. Similar to the location features, we extract: (1) the number of shared time slots when publishing online posts, (2) the inner product of $t(u_i)$ and $t(u_j)$, which is $t(u_i) \cdot t(u_j)$, (3) cosine similarity of these two temporal distribution vectors: $\frac{t(u_i) \cdot t(u_j)}{{\left\|t(u_i)\right\|}\cdot{\left\|t(u_j)\right\|}}$, (4) the Euclidean distance of these two vectors: $(\sum_k{(t(u_i)_k - t(u_j)_k)^2})^{1/2}$ and (5) extend JC in time distribution: $\frac{t(u_i) \cdot t(u_j)}{\left| t(u_i) \right|^2 + \left| t(u_j) \right|^2 - t(u_i) \cdot t(u_j)}.$

 \item \textbf{Text Usage Features}: Text content of posts could can be used as a hint to show the two users' similarity in their word usage habits. In addition, text content could also reveal some personal information of a user. Consider, for example, the fact that a user usually post tweets about food in Twitter strongly implies that the user like food a lot. For this reason, two users with similar word usage habits are also likely to share similar hobbies and are likely to make friends with each other. We transform the words used by two users $u_i$ and $u_j$ into two bag-of-words vectors: $w(u_i)$ and $w(u_j)$ weighted by TF-IDF. Similar features are extracted from these two vectors: (1) the number of shared words used online, (2) inner product of two word vectors: $w(u_i) \cdot w(u_j)$, (3) cosine similarities of these two word vectors: $\frac{w(u_i) \cdot w(u_j)}{{\left\|w(u_i)\right\|}\cdot{\left\|w(u_j)\right\|}}$, (4) the Euclidean distance of these two word vectors: $(\sum_k{(w(u_i)_k - w(u_j)_k)^2})^{1/2}$ and (5) extended JC in text usage: $\frac{w(u_i) \cdot w(u_j)}{\left| w(u_i) \right|^2 + \left| w(u_j) \right|^2 - w(u_i) \cdot w(u_j)}.$
\end{itemize}

\subsection{Link Prediction within Target Network}
After extracting these four categories of features, next we will propose to use them to build supervised methods to predict links for new users in the target network. Before doing that, we notice that the new users' information distribution can be totally different from that of the old users in the target network. However, information of both new users and old users is so important that should be utilized. In this section, we will analyze the differences in information distributions of new users and old users in the target network and propose a personalized within-network sampling method to process old users' information to accommodate the differences. Then, we will revise the traditional supervised link prediction method by using the old users' sampled information in the target network to improve the prediction results.

\subsubsection{Sampling Old Users' Information}
A natural challenge inherent in the usage of the target network to predict social links for new users is the differences in information distributions of new users and old users as mentioned before. To address this problem, we propose a method to accommodate old users' and new users' sub-network by using a within-network personalized sampling method to process old users' information. Totally different from the link prediction with sampling problem studied in \cite{QAH13}, we are conducting personalized sampling within the target network, which contains heterogeneous information, rather across multiple non-aligned homogeneous networks. And the link prediction target are the new users in the target network in our problem.

By sampling the old users' sub-network, we want to meet the following objectives:
\begin{itemize}
	\item \textit{Maximizing Relevance}: We aim at maximizing the relevance of the old users' sub-network and the new users' sub-network to accommodate differences in information distributions of new users and old users in the heterogeneous target network.
	\item \textit{Information Diversity}: Diversity of old users' information after sampling is still of great significance and should be preserved. 
	\item \textit{Structure Maintenance}: Some old users possessing sparse social links should have higher probability to survive after sampling to maintain their links so as to maintain the network structure.
\end{itemize}

Let the heterogeneous target network be $G^t = \{V^t, E^t\}$, and $U^t = U^t_{old} \cup U^t_{new} \subset V^t$ is the set of user nodes (i.e., set of old users and new users) in the target network. Personalized sampling is conducted on the old users' part: $G^t_{old} = \{V^t_{old}, E^t_{old}\}$, in which each node is sampled independently with the sampling rate distribution vector \boldsymbol{$\delta$} = ($\delta_1$, $\delta_2$, $\cdots$, $\delta_n$), where $n = \left| U^t_{old} \right|$, $\sum_{i=1}^n \delta_i = 1$ and $\delta_i \ge 0$. Old users' heterogeneous sub-network after sampling is denoted as $\bar{G}^t_{old} = \{\bar{V}^t_{old}, \bar{E}^t_{old}\}$.

We aim at making the old users' sub-network as relevant to new users' as possible. To measure the similarity score of a user $u_i$ and a heterogeneous network $G$, we define a relevance function as follows:
$$R(u_i, G) = \frac{1}{\left| U \right|}\sum_{u_j \in U}S(u_i, u_j)$$
where set $U$ is the user set of network $G$ and $S(u_i, u_j)$ measures the similarity between user $u_i$ and $u_j$ in the network. Each user has social relationships as well as other auxiliary information and $S(u_i, u_j)$ is defined as the average of similarity scores of these two parts:
$$S(u_i, u_j) = \frac{1}{2}(S_{aux}(u_i, u_j) + S_{social}(u_i, u_j))$$

In our problem settings, the auxiliary information of each users could also be divided into 3 categories: \textit{location}, \textit{temporal}, and \textit{text}. So, $S_{aux}(u_i, u_j)$ is defined as the mean of these three aspects. 
$$S_{aux}(u_i, u_j) = \frac{1}{3}(S_{text}(u_i, u_j) + S_{loc}(u_i, u_j) + S_{temp}(u_i, u_j))$$

There are many different methods measuring the similarities of these auxiliary information in different aspects, e.g. cosine similarity. As to the social similarity, Jaccard's Coefficient can be used to depict how similar two users are in their social relationships. 

The relevance between the sampled old users' network and the new users' network could be defined as the expectation value of function $R(\bar{u}^t_{old}, G^t_{new})$:
\begin{align*}
R(\bar{G}^t_{old}, G^t_{new}) &= \mathbb{E}(R(\bar{u}^t_{old}, G^t_{new})) \\
			&= \frac{1}{\left| U^t_{new} \right|} \sum_{j=1}^{\left | U^t_{new} \right |} \mathbb{E}(S(\bar{u}^t_{old}, u^t_{new,j})) \\
			&= \frac{1}{\left| U^t_{new} \right|} \sum_{j=1}^{\left | U^t_{new} \right |} \sum_{i=1}^{\left | U^t_{old} \right |} \delta_i \cdot S(\bar{u}^t_{old,i},u^t_{new,j}) \\
			&= \boldsymbol{\delta'}\boldsymbol{s}
\end{align*}
where vector $\boldsymbol{s}$ equals:
$$\frac{1}{\left| U^t_{new} \right|}[\sum_{j=1}^{\left | U^t_{new} \right |} S(\bar{u}^t_{old,1},u^t_{new,j}), \cdots, \sum_{j=1}^{\left | U^t_{new} \right |} S(\bar{u}^t_{old,n},u^t_{new,j})]^T$$
and ${\left | U^t_{old} \right |} = n$. Besides the relevance, we also need to ensure that the diversity of information in the sampled old users' sub-network could be preserved. Similarly, it also includes diversities of the auxiliary information and social relationships. The diversity of auxiliary information is determined by the sampling rate $\delta_i$, which could be define with the averaged \textit{Simpson Index} \cite{S49} over the old users' sub-network.
$$D_{aux}(\bar{G}^t_{old}) = \frac{1}{\left| U^t_{old} \right|} \cdot \sum_{i=1}^{\left| U^t_{old} \right|}{\delta_i^2}$$
As to the diversity in the social relationship, we could get the existence probability of a certain social link $(u_i, u_j)$ after sampling to be proportional to $\delta_i \cdot \delta_j$. So, the diversity of social links in the sampled network could be defined as average existence probabilities of all the links in the old users' sub-network.
$$D_{social}(\bar{G}^t_{old}) =\frac{1}{{\left| S^t_{old} \right|}} \cdot \sum_{i=1}^{\left| U^t_{old} \right|}\sum_{j=1}^{\left| U^t_{old} \right|}\delta_i \cdot \delta_j \times I(u_i, u_j)$$
where $\left| S^t_{old} \right|$ is the size of social link set of old users' sub-network and  $I(u_i, u_j)$ is an indicator function $I: (u_i, u_j) \to \{0, 1\}$ to show whether a certain social link exists or not originally before sampling. For example, if link $(u_i, u_j)$ is a social link in the target network originally before sampling, then $I(u_i, u_j) = 1$, otherwise it equals to $0$.
%

Considering these two terms simultaneously, we could have the diversity of information in the sampled old users' sub-network to be the average diversities of these two parts:

\begin{align*}
D(\bar{G}^t_{old}) &= \frac{1}{2}(D_{social}(\bar{G}^t_{old}) + D_{aux}(\bar{G}^t_{old}))\\
			&= \frac{1}{2}(\sum_{i=1}^{\left| U^t_{old} \right|}\sum_{j=1}^{\left| U^t_{old} \right|}\frac{1}{{\left| S^t_{old} \right|}} \cdot \delta_i \cdot \delta_j \times I(u_i, u_j)\\
			&+ \sum_{i=1}^{\left| U^t_{old} \right|}\frac{1}{\left| U^t_{old} \right|} \cdot {\delta_i^2} )\\
			&= \boldsymbol{\delta}' \cdot (\frac{1}{2{\left| S^t_{old} \right|}} \cdot \mathbf{A_{old}^t} + \frac{1}{2\left| U^t_{old} \right|} \cdot \mathbf{I_{\left| U^t_{old} \right|}}) \cdot \boldsymbol{\delta}
\end{align*}
where matrix $\mathbf{I_{\left| U^t_{old} \right|}}$ is the diagonal identity matrix of size ${\left| U^t_{old} \right|} \times {\left| U^t_{old} \right|}$ and $\mathbf{A_{old}^t}$ is the adjacency matrix of old users' sub-network. 

To ensure that the structure of the original old users' subnetwork is not destroyed, we need to ensure that users with few links could also preserve their links. So, we could add a regularization term to increase the sampling rate for these users as well as their neighbours by maximizing the following terms:
\begin{align*}
	Reg(\bar{G}^t_{old}) &= \min\{\mathcal{N}_i, \min_{u_j \in \mathcal{N}_i}\{\mathcal{N}_j\}\} \times \delta_i^2 = \boldsymbol{\delta}' \cdot \mathbf{M} \cdot \boldsymbol{\delta}
\end{align*}
where matrix $\mathbf{M}$ is a diagonal matrix with element $\mathbf{M}_{i,i} = \min\{\mathcal{N}_i, \min_{u_j \in \mathcal{N}_i}\{\mathcal{N}_j\}\} = \min\{\mathcal{N}_i, \{\mathcal{N}_i| u_j \in \mathcal{N}_i\}$ and $\mathcal{N}_j = |\Gamma(u_j)|$ is the size of user $u_j$'s neighbour set. So, if a user or his/her neighbours have few links, then this user as well as his/her neighbours should have higher sampling rate so as to preserve the links between them.

\begin{figure}[!t]
 \centering    
 \begin{minipage}[l]{0.8\columnwidth}
  \centering
    \includegraphics[width=1.0\textwidth]{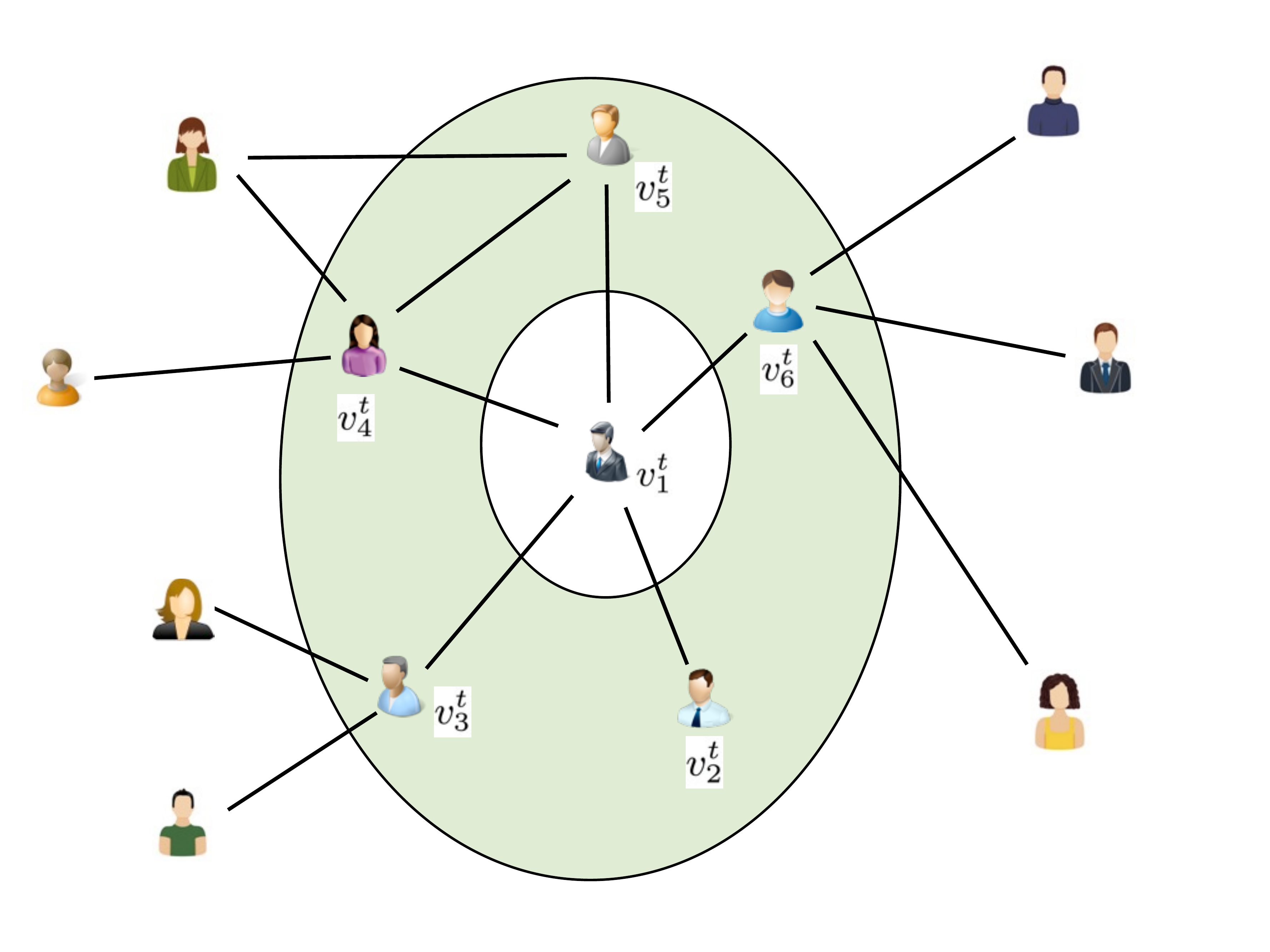}
 \end{minipage}
\caption{Personalized sampling preserving network structures.}\label{fig_eg4}
\end{figure}

\begin{figure*}[t]
\centering
\subfigure[{\ourtrad} method]{ \label{fig_eg3_1}
   \begin{minipage}[l]{.6\columnwidth}
       \centering
      \includegraphics[width=1.0\textwidth]{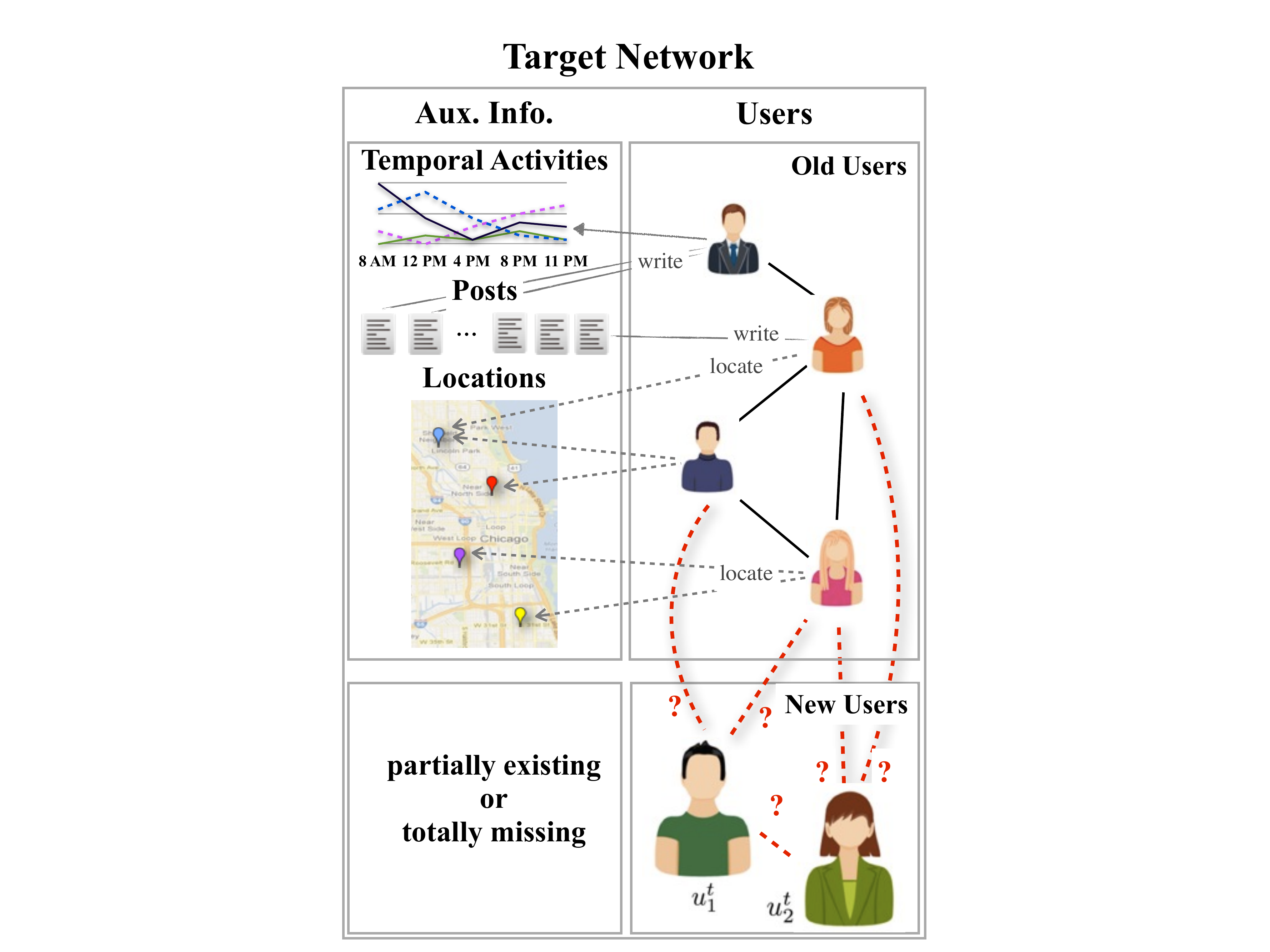}
    \end{minipage}
}
\subfigure[{\ournaive} method]{ \label{fig_eg3_2}
   \begin{minipage}[l]{.6\columnwidth}
       \centering
      \includegraphics[width=1.0\textwidth]{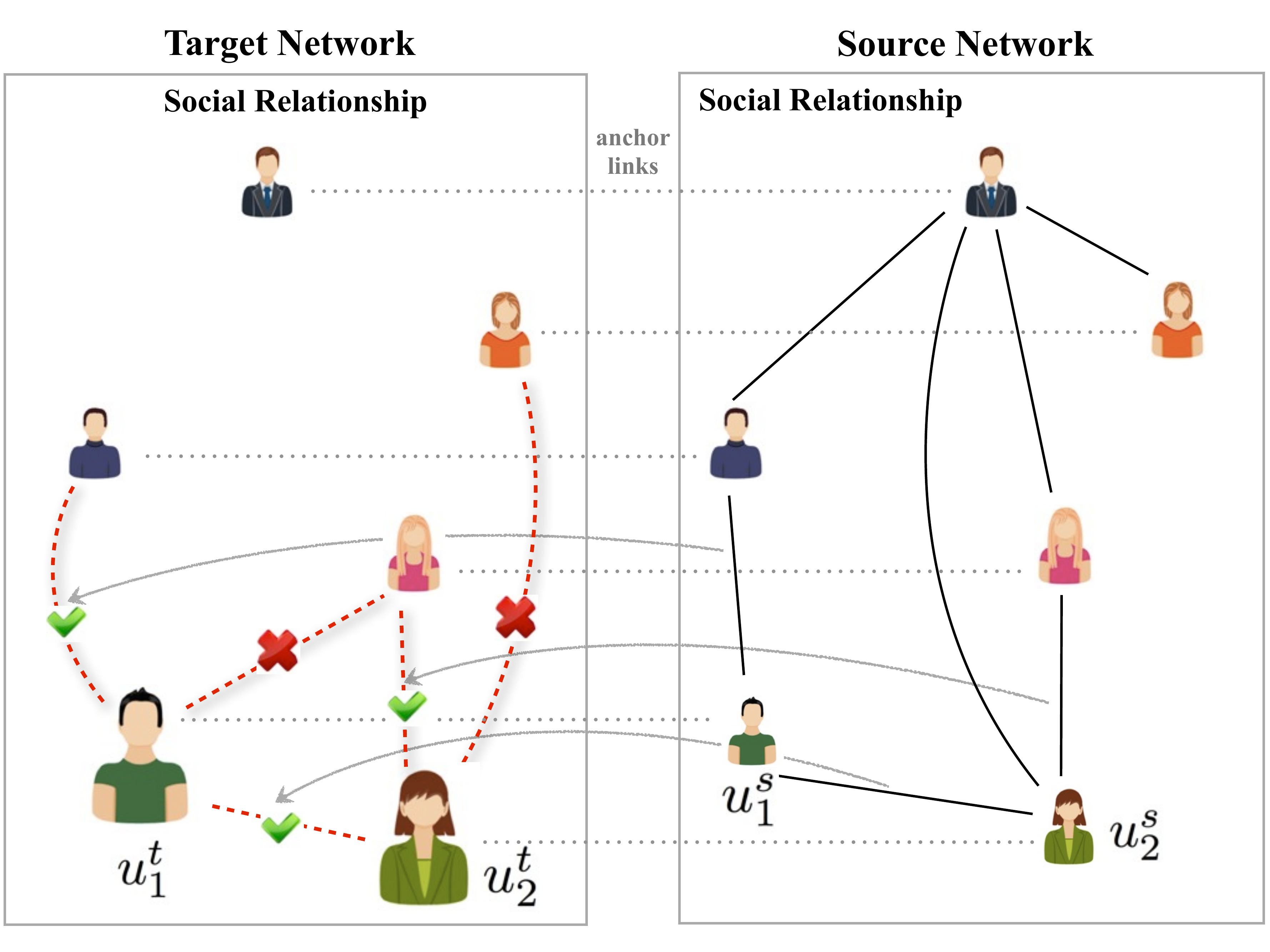}
    \end{minipage}
}
\subfigure[{\our} method]{ \label{fig_eg3_3}
   \begin{minipage}[l]{.6\columnwidth}
       \centering
      \includegraphics[width=1.0\textwidth]{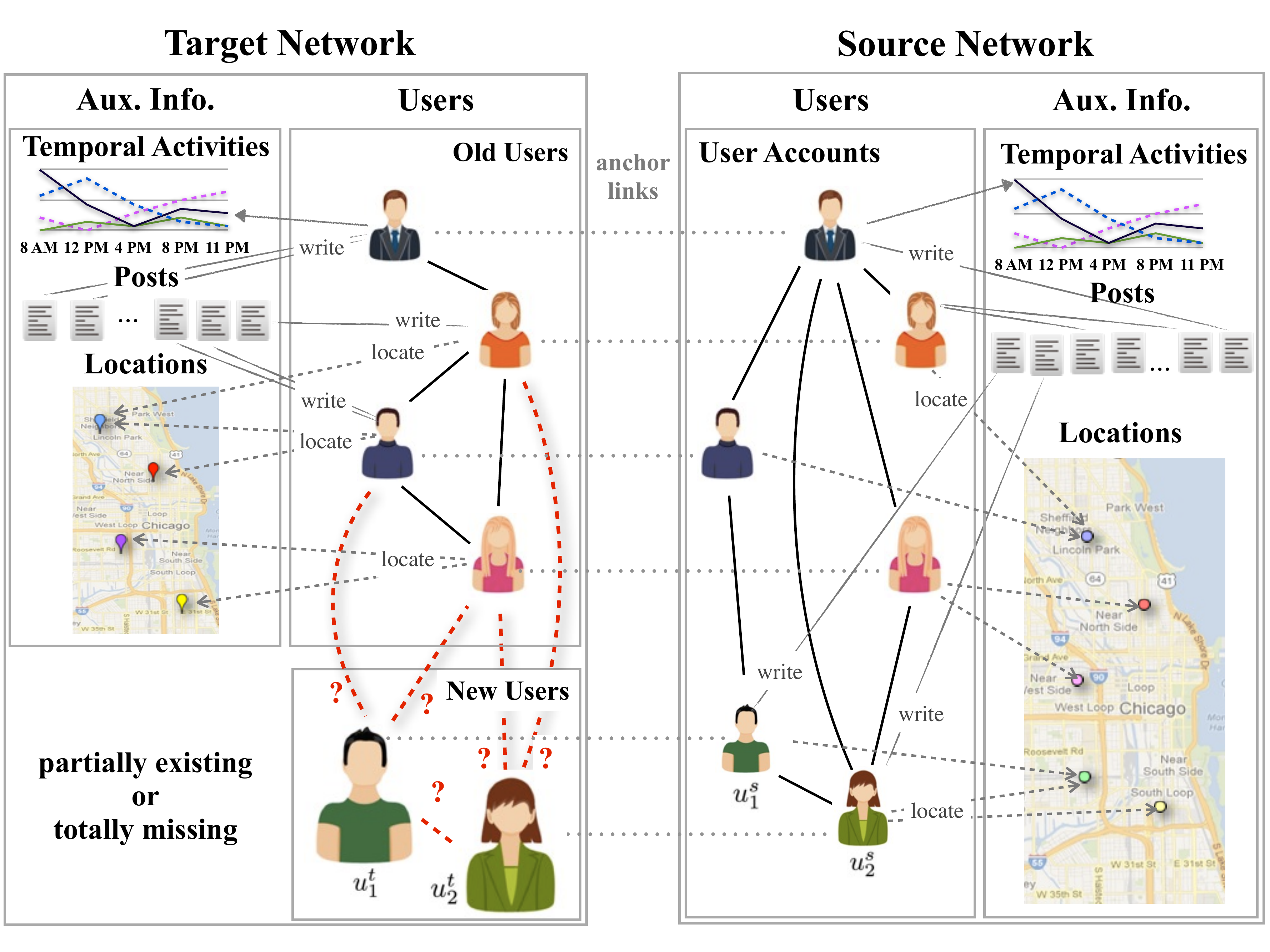}
    \end{minipage}
}

\caption{Different methods to predict social link for new users.}\label{fig_eg3}
\end{figure*}

For example, in Figure~\ref{fig_eg4}, we have $6$ users. To decide the sampling rate of user $u^t_1$, we need to consider his/her social structure. We find that since $u^t_1$'s neighbour $u^t_2$ has no other neighbour except $u^t_1$. To preserve the social link between $u^t_1$ and $u^t_2$ we need to increase the sampling rate of $u^t_2$. However, the existence probability of link $(u^t_1, u^t_2)$ is also decided by the sampling rate of user $u^t_1$, which also needs to be increased too.

Combining the diversity term and the structure preservation term, we could define the regularized diversity of information after sampling to be
\begin{align*}
D_{Reg}(\bar{G}^t_{old}) &= D(\bar{G}^t_{old}) + Reg(\bar{G}^t_{old}) = \boldsymbol{\delta}' \cdot \mathbf{N} \cdot \boldsymbol{\delta}
\end{align*}
where $\mathbf{N} = \mathbf{ \frac{1}{2\left| U^t_{old} \right|} \cdot I_{\left| U^t_{old} \right|}} + \mathbf{\frac{1}{{2\left| S^t_{old} \right|}} \cdot \mathbf{A_{old}^t}}+  \mathbf{M}$.

The optimal value of $\boldsymbol{\delta}$ should be able to maximize the relevance of new users' sub-network and old users' as well as the regularized diversity of old users' information in the target network
\begin{align*}
\boldsymbol{\delta} &= \underset{\boldsymbol{\delta}}{\arg \max}\  R(\bar{G}^t_{old}, G^t_{new}) + \theta \cdot D_{Reg}(\bar{G}^t_{old}) \\
	&= \underset{\boldsymbol{\delta}}{\arg \max}\  \boldsymbol{\delta'}\boldsymbol{s} + \theta \cdot \boldsymbol{\delta}' \cdot \mathbf{N} \cdot \boldsymbol{\delta}\\
	&s.t., \sum_{i=1}^{\left| U^t_{old} \right|}\delta_i = 1\ and\ \delta_i \ge 0.
\end{align*}
where, parameter $\theta$ is used to weight the importance of term regularized information diversity.
\subsubsection{\ourtrad} 
A traditional supervised link prediction method {\ourtrad} (Traditional Link Prediction) for our task is to use the existing links in the target network to train a classifier and apply it to classify the potential social links for new users. In method {\ourtrad}, only the target network is used, which consists of new users and unsampled old users. To overcome the difference in information distribution between new users and old users in the target network, we revise it a little and get method: {\ourtradsp} (Traditional Link Prediction with Personalized Sampling). {\ourtradsp} consists of two steps: (1) personalized sampling of the old users' sub-network with the previous method; (2) usage of similar techniques as {\ourtrad} to predict links based on the sampled network. Theoretically, {\ourtrad} and {\ourtradsp} could work well by using information in the target network. However, considering the fact it is impossible for new users to possess large amount of information actually, {\ourtrad} and {\ourtradsp} would suffer from the long-standing cold start problem caused by the lack of historical information indicating these new users' preferences. This problem will be even worse when dealing with brand-new users, who have no information at all in the target network. 

For example, in Figure~\ref{fig_eg3_1}, user $u^t_1$ and $u^t_2$ are two new users in the target network, who possess very few social links with other users and little auxiliary information. We can not get any information about these two new users and the information we could use is that possessed by other old users. As a result, the links that {\ourtrad} and {\ourtradsp} predicted could hardly be of high quality.

In order to deal with such problem, we will use aligned networks simultaneously in the next section.

\subsection{Cold-Start Link Prediction}
In our problem settings, we have two aligned social networks and the methods proposed in previous section using the target network may suffer from the cold start problems when processing brand-new users. In this section, we will propose two methods to utilize the aligned source network to help solve the problem and improve the prediction results.
\subsubsection{\ournaive}
Suppose we have a new user $u_i^t$ in the target network, a naive way to use the aligned source network to recommend social links for user $u^t_i$ is to recommend all the corresponding social links related to this user's aligned account $u^s_i$ in the aligned source network to him/her. Based on this intuition, we propose a cold start link prediction method {\ournaive} (Naive Link Prediction). To clarify how {\ournaive} works in reality, we will give an example next. And before that, we formally define a new term \textit{pseudo label} to denote the existence of corresponding links in the aligned source network.

\noindent \textbf{Definition 4} (Pseudo Label): The pseudo label of a link $(u^t_i, u^t_j)$ in the target denotes the existence of its corresponding link $(u^s_i, u^s_j)$ in the aligned source network and it is $1$ if $(u^s_i, u^s_j)$ exists and $0$ otherwise.

For example, in Figure~\ref{fig_eg3_2},  to decide whether to recommend $u^t_1$ to $u^t_2$ in the target network or not, we could find their aligned accounts: $u^s_1$ and $u^s_2$, and their social link: $(u^s_1, u^s_2)$ in the aligned source network with the help of \textit{anchor links}. We find that $u^s_1$ and $u^s_2$ are friends in the aligned source network and link $(u^s_1, u^s_2)$ exists in the aligned source network. As a result, the pseudo label of link $(u^t_1, u^t_2)$ is $1$ and in the target network, we could recommend $u^t_2$ to $u^t_1$. And that is the reason why the social link between $u^t_1$ and $u^t_2$ is predicted to be existent by method {\ournaive}. Other links in Figure~\ref{fig_eg3_2} are predicted in a similar way.

Method {\ournaive} is very simple and could work well in our task even when these new users are brand new, which means that we could overcome the cold start problem by using this method. However, it may still suffer from some disadvantages: (1) the social structures of different networks are not always identical which will degrade the performance of {\ournaive} a lot; (2) {\ournaive}  only utilizes these new users' social linkage information in the source network and ignores all other information.

\subsubsection{\our}
To overcome all these disadvantages mentioned above, a new method {\our} (Supervised Cross Aligned Networks Link Prediction with Personalized Sampling) is proposed. As shown in Figure~\ref{fig_eg3_3}, it could use heterogeneous information existing in both the target network and the aligned source and it is built across two aligned social networks. Taking advantage of the anchor links, we could locate the users' aligned accounts and their information in the aligned source network exactly. If two aligned networks are used simultaneously, different categories of features are extracted from aligned networks. To use multiple networks, these feature vectors extracted for the corresponding links in aligned networks are merged into an expanded feature vector. The expanded feature vector together with the labels from the target network are used to build a cross-network classifier to decide the existence of social links related to these new users in the target network. This is how method {\our} works. {\our} is quite stable and could overcome the cold start problem for the reason that the information about all these users in the aligned source network doesn't change much with the variation of the target network and we get the information showing of these new users' preferences from the information he/she leaves in the aligned source network. As the old users' information inside the target network is also used in {\our}, personalized sampling is also conducted to preprocess the old users' information in the target network.


In addition to features mentioned before, method {\our} also utilizes the information used by {\ournaive}, which is the \textit{pseudo label} defined before, by regarding it as an extra feature. 
\begin{itemize}
\item \textbf{An Extra Feature}: We use the pseudo label as an extract feature to denote the existence of the corresponding links in the aligned source network. 
\end{itemize}

Compared with {\our} with {\ournaive}, {\our} has many advantages: (1) {\our} utilizes multiple categories of information; (2) {\our} can make use of the information hidden in the old users' network by incorporating them into the training set; (3) {\our} doesn't rely on the assumption that the social relationships in different networks are identical, which is very risky actually.

Compared with {\ourtrad} and {\ourtradsp},  {\our} can solve the cold start problem as it could have access to information owned by these new users in other aligned source networks. Similar to {\ourtrad} and {\ourtradsp}, these new users' information is used if they are not very new and other old users' information in the target is also preprocessed by using the within-network personalized sampling method before the intra-network knowledge transfer.

\begin{table}[t]
\caption{Properties of the Heterogeneous Social Networks}
\label{tab:datastat}
\centering
\begin{tabular}{clrr}
\toprule
&&\multicolumn{2}{c}{network}\\
\cmidrule{3-4}
&property &\textbf{Twitter}	&\textbf{Foursquare}	 \\
\midrule 
\multirow{3}{*}{\# node}
&user		& 5,223	& 5,392 \\
&tweet/tip	& 9,490,707	& 48,756 \\
&location	& 297,182	& 38,921 \\
\midrule 
\multirow{3}{*}{\# link}
&friend/follow		&164,920	&31,312 \\
&write		& 9,490,707	& 48,756 \\
&locate		& 615,515	& 48,756 \\
\bottomrule
\end{tabular}
\end{table}

\section{Experiments}\label{sec:experiment}
\subsection{Data Preparation}
In previous sections, we propose method {\our} to solve the challenges mentioned in the introduction section. To test whether our method is applicable in reality, we conduct extensive experiments on two real-world aligned heterogeneous social networks: Twitter and Foursquare. A more detailed description about these two social networks datasets is summarized in Table~\ref{tab:datastat}. We crawl these two datasets by using the APIs provided by Twitter and Foursquare.

\begin{itemize}
  \item \textbf{Twitter}: The first social network used by us is Twitter, a famous online microblogging network containing lots of heterogeneous information. 5,223 users together with 9,490,707 tweets posted by them are crawled. Each tweet could contain the text content, timestamps and locations. About 615,515 tweets are found to possess location information(latitude and longitude), which accounts for about 6.5\% of all the tweets.
  \item \textbf{Foursquare}: Another social network used by us is Foursquare, which is a well-known location-based social network(LBSN). We crawled 5,392 users and their tips, the number of which is 48,756. Similar to tweets, content information, timestamps as well as the specific locations(latitude and longitude) are also available. Different from Twitter, in Foursquare, each tip is related to a location and the number of locate links is equal to the number of tips.
\end{itemize}

Both of these two social networks contain social links, which are used as the ground truth in the experiments. The anchor links between these two networks is acquired by crawling the hyperlink of the users' Twitter account in their Foursquare homepages.

\begin{table*}[t]
\caption{Performance comparison of different social link prediction methods for users of different degrees of newness. Target network: Foursquare. Source network: Twitter. (Degree of newness denotes the ratio of information owned by users)} 
\label{tab:setting1}
\centering
{\scriptsize
\begin{tabular}{lrccccccccccc}
\toprule
& &\multicolumn{8}{c}{\textsc{Remaining Information Ratio}}\\
\cmidrule{3-10}
\multicolumn{1}{l}{measure}	&\multicolumn{1}{r}{method}	&0.0	& 0.1	& 0.2	& 0.3	&0.4	&0.5	&0.6	& 0.7 \\ 
\midrule
\multirow{9}{*}{\textsc{Auc}}

&{\our}	&\textbf{0.783}$\pm$\textbf{0.009}	&\textbf{0.839}$\pm$\textbf{0.008}	&\textbf{0.864}$\pm$\textbf{0.013}	&\textbf{0.883}$\pm$\textbf{0.008}	&\textbf{0.902}$\pm$\textbf{0.011}	&\textbf{0.910}$\pm$\textbf{0.009}	&\textbf{0.912}$\pm$\textbf{0.003}	&\textbf{0.913}$\pm$\textbf{0.012}	\\
&{\ourb}	&0.768$\pm$0.013	&0.808$\pm$0.007	&0.833$\pm$0.009	&0.846$\pm$0.006	&0.854$\pm$0.005	&0.860$\pm$0.008	&0.869$\pm$0.009	&0.882$\pm$0.006	\\
&{\ourd}	&0.761$\pm$0.008	&0.768$\pm$0.015	&0.800$\pm$0.014	&0.802$\pm$0.011	&0.806$\pm$0.003	&0.815$\pm$0.011	&0.820$\pm$0.006	&0.820$\pm$0.007	\\
&{\ourtradsp}	&0.553$\pm$0.007	&0.626$\pm$0.003	&0.69$\pm$0.012	&0.681$\pm$0.012	&0.701$\pm$0.008	&0.701$\pm$0.007	&0.735$\pm$0.014	&0.736$\pm$0.013\\	
&{\ourf}	&0.554$\pm$0.016	&0.567$\pm$0.01	&0.564$\pm$0.022	&0.571$\pm$0.012	&0.558$\pm$0.005	&0.578$\pm$0.009	&0.570$\pm$0.015	&0.575$\pm$0.010	\\
&{\ourtrad}	&0.555$\pm$0.006	&0.593$\pm$0.007	&0.622$\pm$0.009	&0.646$\pm$0.012	&0.658$\pm$0.006	&0.671$\pm$0.016	&0.681$\pm$0.010	&0.708$\pm$0.011	\\
&{\ourh}	&0.550$\pm$0.008	&0.510$\pm$0.010	&0.527$\pm$0.008	&0.541$\pm$0.015	&0.551$\pm$0.006	&0.571$\pm$0.012	&0.574$\pm$0.010	&0.568$\pm$0.009	\\
&{\ouri}	&0.495$\pm$0.018	&0.616$\pm$0.011	&0.631$\pm$0.005	&0.646$\pm$0.006	&0.653$\pm$0.009	&0.656$\pm$0.004	&0.670$\pm$0.010	&0.675$\pm$0.009	\\

\cmidrule{3-10}
&CN	&0.500$\pm$0.000	&0.523$\pm$0.005	&0.536$\pm$0.004	&0.552$\pm$0.006	&0.562$\pm$0.004	&0.573$\pm$0.005	&0.576$\pm$0.007	&0.587$\pm$0.003 \\ 
&JC	&0.500$\pm$0.000	&0.523$\pm$0.005	&0.534$\pm$0.006	&0.554$\pm$0.007	&0.562$\pm$0.010	&0.572$\pm$0.005	&0.575$\pm$0.009	&0.587$\pm$0.004 \\ 
&AA	&0.500$\pm$0.000	&0.521$\pm$0.004	&0.531$\pm$0.003	&0.548$\pm$0.006	&0.556$\pm$0.004	&0.566$\pm$0.004	&0.569$\pm$0.006	&0.583$\pm$0.002 \\ 
\midrule
\multirow{9}{*}{Acc.}

&{\our}	&\textbf{0.747}$\pm$\textbf{0.005}	&\textbf{0.772}$\pm$\textbf{0.010}	&\textbf{0.802}$\pm$\textbf{0.007}	&\textbf{0.811}$\pm$\textbf{0.009}	&\textbf{0.813}$\pm$\textbf{0.012}	&\textbf{0.821}$\pm$\textbf{0.008}	&\textbf{0.826}$\pm$\textbf{0.005}	&\textbf{0.834}$\pm$\textbf{0.008}	\\
&{\ourb}	&0.732$\pm$0.014	&0.746$\pm$0.008	&0.763$\pm$0.010	&0.778$\pm$0.007	&0.791$\pm$0.008	&0.790$\pm$0.009	&0.794$\pm$0.009	&0.803$\pm$0.009	\\
&{\ourd}	&0.695$\pm$0.011	&0.712$\pm$0.011	&0.716$\pm$0.015	&0.733$\pm$0.009	&0.738$\pm$0.003	&0.735$\pm$0.012	&0.745$\pm$0.009	&0.740$\pm$0.006	\\
&{\ourtradsp}	&0.506$\pm$0.004	&0.600$\pm$0.006	&0.610$\pm$0.009	&0.625$\pm$0.005	&0.628$\pm$0.005	&0.632$\pm$0.009	&0.645$\pm$0.006	&0.653$\pm$0.007	\\
&{\ourf}	&0.506$\pm$0.002	&0.504$\pm$0.002	&0.505$\pm$0.004	&0.512$\pm$0.026	&0.518$\pm$0.006	&0.535$\pm$0.010	&0.520$\pm$0.015	&0.524$\pm$0.026	\\
&{\ourtrad}	&0.506$\pm$0.002	&0.524$\pm$0.006	&0.540$\pm$0.004	&0.559$\pm$0.006	&0.586$\pm$0.009	&0.599$\pm$0.007	&0.624$\pm$0.012	&0.635$\pm$0.009	\\
&{\ourh}	&0.503$\pm$0.002	&0.503$\pm$0.002	&0.503$\pm$0.004	&0.505$\pm$0.003	&0.505$\pm$0.003	&0.515$\pm$0.004	&0.509$\pm$0.005	&0.516$\pm$0.003	\\
&{\ouri}	&0.478$\pm$0.010	&0.563$\pm$0.009	&0.581$\pm$0.004	&0.591$\pm$0.007	&0.602$\pm$0.009	&0.604$\pm$0.006	&0.615$\pm$0.010	&0.628$\pm$0.005	\\

\cmidrule{3-10}
&{\ournaive}	&0.616$\pm$0.009	&0.608$\pm$0.004	&0.622$\pm$0.003	&0.616$\pm$0.008	&0.619$\pm$0.009	&0.613$\pm$0.003	&0.615$\pm$0.009	&0.614$\pm$0.008 \\
\bottomrule
\end{tabular}
}
\end{table*}

\begin{table*}[t]

\caption{Performance comparison of different social link prediction methods for users of different degrees of newness. Target network: Twitter. Source network: Foursquare. (Degree of newness denotes the ratio of information owned by users)} 
\label{tab:setting2}
\centering
{\scriptsize
\begin{tabular}{lrccccccccccc}
\toprule
&	&\multicolumn{8}{c}{\textsc{Remaining Information Ratio}}\\
\cmidrule{3-10}
\multicolumn{1}{l}{measure}	&\multicolumn{1}{r}{method}	&0.0	& 0.1	& 0.2	& 0.3	&0.4	&0.5	&0.6	& 0.7 \\ 
\midrule
\multirow{9}{*}{\textsc{Auc}}

&{\our}	&\textbf{0.608}$\pm$\textbf{0.006}	&\textbf{0.832}$\pm$\textbf{0.005}	&\textbf{0.859}$\pm$\textbf{0.004}	&\textbf{0.886}$\pm$\textbf{0.003}	&\textbf{0.890}$\pm$\textbf{0.003}	&\textbf{0.899}$\pm$\textbf{0.004}	&\textbf{0.911}$\pm$\textbf{0.005}	&\textbf{0.910}$\pm$\textbf{0.005}	\\
&{\ourb}	&0.602$\pm$0.005	&0.788$\pm$0.005	&0.827$\pm$0.003	&0.851$\pm$0.005	&0.850$\pm$0.007	&0.854$\pm$0.003	&0.870$\pm$0.004	&0.884$\pm$0.002	\\
&{\ourd}	&0.621$\pm$0.007	&0.736$\pm$0.005	&0.734$\pm$0.005	&0.743$\pm$0.006	&0.745$\pm$0.004	&0.743$\pm$0.001	&0.749$\pm$0.003	&0.749$\pm$0.008	\\
&{\ourtradsp}	&0.526$\pm$0.004	&0.772$\pm$0.006	&0.785$\pm$0.002	&0.807$\pm$0.006	&0.822$\pm$0.005	&0.837$\pm$0.002	&0.841$\pm$0.003	&0.857$\pm$0.004	\\
&{\ourf}	&0.530$\pm$0.003	&0.680$\pm$0.007	&0.653$\pm$0.006	&0.644$\pm$0.007	&0.635$\pm$0.007	&0.640$\pm$0.002	&0.627$\pm$0.004	&0.542$\pm$0.010	\\
&{\ourtrad}	&0.456$\pm$0.003	&0.697$\pm$0.007	&0.772$\pm$0.004	&0.801$\pm$0.006	&0.820$\pm$0.004	&0.833$\pm$0.005	&0.846$\pm$0.005	&0.858$\pm$0.004	\\
&{\ourh}	&0.423$\pm$0.002	&0.519$\pm$0.004	&0.528$\pm$0.005	&0.568$\pm$0.006	&0.600$\pm$0.006	&0.629$\pm$0.003	&0.653$\pm$0.003	&0.674$\pm$0.004	\\
&{\ouri}	&0.492$\pm$0.013	&0.766$\pm$0.008	&0.788$\pm$0.003	&0.806$\pm$0.005	&0.822$\pm$0.004	&0.834$\pm$0.004	&0.842$\pm$0.005	&0.851$\pm$0.004	\\

\cmidrule{3-10}
&CN	&0.500$\pm$0.000	&0.731$\pm$0.006	&0.786$\pm$0.001	&0.814$\pm$0.006	&0.821$\pm$0.005	&0.830$\pm$0.005	&0.837$\pm$0.003	&0.839$\pm$0.003 \\ 
&JC	&0.500$\pm$0.000	&0.716$\pm$0.007	&0.760$\pm$0.002	&0.789$\pm$0.006	&0.794$\pm$0.006	&0.804$\pm$0.007	&0.810$\pm$0.003	&0.813$\pm$0.003 \\ 
&AA	&0.500$\pm$0.000	&0.728$\pm$0.005	&0.782$\pm$0.002	&0.811$\pm$0.004	&0.818$\pm$0.005	&0.828$\pm$0.007	&0.835$\pm$0.003	&0.837$\pm$0.003 \\ 
\midrule
\multirow{9}{*}{Acc.}

&{\our}	&\textbf{0.588}$\pm$\textbf{0.001}	&\textbf{0.769}$\pm$\textbf{0.004}	&\textbf{0.793}$\pm$\textbf{0.005}	&\textbf{0.815}$\pm$\textbf{0.004}	&\textbf{0.822}$\pm$\textbf{0.002}	&\textbf{0.848}$\pm$\textbf{0.004}	&\textbf{0.860}$\pm$\textbf{0.005}	&\textbf{0.868}$\pm$\textbf{0.004}\\
&{\ourb}	&0.582$\pm$0.004	&0.685$\pm$0.007	&0.715$\pm$0.004	&0.731$\pm$0.004	&0.753$\pm$0.008	&0.776$\pm$0.004	&0.791$\pm$0.004	&0.817$\pm$0.003	\\
&{\ourd}	&0.573$\pm$0.006	&0.669$\pm$0.005	&0.676$\pm$0.003	&0.680$\pm$0.005	&0.684$\pm$0.002	&0.683$\pm$0.004	&0.686$\pm$0.003	&0.686$\pm$0.008\\
&{\ourtradsp}	&0.505$\pm$0.002	&0.710$\pm$0.001	&0.705$\pm$0.005	&0.741$\pm$0.006	&0.753$\pm$0.005	&0.765$\pm$0.003	&0.769$\pm$0.003	&0.778$\pm$0.005\\
&{\ourf}	&0.515$\pm$0.003	&0.501$\pm$0.013	&0.503$\pm$0.002	&0.502$\pm$0.010	&0.512$\pm$0.002	&0.502$\pm$0.002	&0.503$\pm$0.052	&0.501$\pm$0.003	\\
&{\ourtrad}	&0.503$\pm$0.002	&0.545$\pm$0.005	&0.625$\pm$0.002	&0.680$\pm$0.009	&0.723$\pm$0.002	&0.745$\pm$0.003	&0.763$\pm$0.004	&0.767$\pm$0.005\\
&{\ourh}	&0.516$\pm$0.006	&0.500$\pm$0.002	&0.513$\pm$0.001	&0.504$\pm$0.002	&0.503$\pm$0.002	&0.510$\pm$0.002	&0.500$\pm$0.001	&0.503$\pm$0.002\\
&{\ouri}	&0.488$\pm$0.008	&0.661$\pm$0.006	&0.707$\pm$0.003	&0.731$\pm$0.004	&0.743$\pm$0.004	&0.758$\pm$0.005	&0.765$\pm$0.003	&0.775$\pm$0.004	\\

\cmidrule{3-10}
&{\ournaive}	&0.552$\pm$0.003	&0.552$\pm$0.002	&0.553$\pm$0.002	&0.552$\pm$0.004	&0.554$\pm$0.003	&0.553$\pm$0.004	&0.553$\pm$0.002	&0.552$\pm$0.003 \\
\bottomrule
\end{tabular}
}
\end{table*}


\subsection{Experiment Settings}
\textbf{Comparison Methods}: To evaluate the effectiveness of {\our} in predicting social links for new users, we compare {\our} with many baseline methods, including both supervised and unsupervised methods. To ensure the fairness of the comparisons, LibSVM \cite{libsvm} of linear kernel with default parameter is used as the base classifier for all supervised methods. The evaluation methods used by us are: AUC and Accuracy. Next, we will summary all the comparative methods first and then give the description of the experiment settings and the evaluation method.
\begin{itemize}
 \item \textit{Source Network + Target Network}: {\our} could use the information in the aligned source network and the target network at the same time. Old users' information in the target network is used by {\our} and it is processed with personalized sampling method before being transferred. To show personalized sampling of old users' information is helpful for our task, we compare it with another weaker baseline method, which use old users' information without sampling. The method is named as {\ourb} (Supervised Cross Aligned Networks Link Prediction).

 \item \textit{Target Network Only}: Some other supervised baseline methods are built only with the information in target network. Method {\ouri} is built only with new users' information, while method {\ourh} only uses the information old users' information in the target network. Method {\ourtrad} could use all the information in the target network. To show that personalized sampling of old users' information is helpful, methods {\ourtradsp} and {\ourf} are used to compare with {\ourtrad} and {\ourh} respectively.

 \item \textit{Source Network Only}: To show that using two networks simultaneously is better than using one network. Besides those methods using target network only, we also compare {\our} with another baseline method {\ourd}, built with all the information in the target network. 

 \item \textit{Unsupervised Methods}: {\ournaive} and some traditional unsupervised social link prediction methods are also used as the unsupervised baseline methods to be compared with {\our}. The other unsupervised baseline methods include \textit{Common Neighbour} (CN), \textit{Jaccard Coefficient} (JC) and \textit{Adamic Adar} (AA). {\ournaive} uses social information in the aligned source network only, while all other three methods are based on the target network without sampling.
\end{itemize}

\textbf{Experiment Setting}: To get two fully aligned networks, 1000 users in each of these two networks with full anchor links are randomly sampled with breadth-first-search and these users' complete social links and other auxiliary information is preserved. Then, we randomly sample 20\% of these 1000 users in the target network as new users and the remaining are regarded as old users. All the social links related to new users are grouped into a positive link set and equivalent number of non-existent social links related these new users are organized into a negative link set. We partition these two link sets into two groups by 5-fold cross validation: four folds are used as the training set and the remaining one fold is used as the testing set. To get different degree of newness, all the information, i.e., social links and other auxiliary information, owned by these new users inside the network are randomly sample with a certain rate denoting the novelty. If the old users' information is used, we use the within-network personalized sampling method to preprocess the old users' information inside the target network before the intra-network transfer. The personalized sampling vector \boldsymbol{$\delta$} is learnt from the target network. All the existent social links related to the old users after sampling and equivalent number of nonexistent links in the target network are added to the training set. Heterogeneous features of each positive and negative link are extracted from the aligned networks. If two networks are used simultaneously, the feature vectors extracted from each social network are merged into expanded ones. There are two networks in our dataset and we choose Foursquare as the target network and Twitter as the source network first. And, then use them in a reverse way. 

\textbf{Evaluation Methods}: Evaluation methods utilized by us are \textsc{Auc} and Accuracy. Since the three unsupervised methods $CN, JC, AA$ could only predict a real-number score to measure the confidence about the existence of a certain social link, we only use \textsc{Auc} to evaluate these methods' performance. And {\ournaive} could only predict the labels without confidence, so it is evaluated only by Accuracy. All other methods are evaluated by both \textsc{Auc} and Accuracy.

\subsection{Experiment Result}
In Table~\ref{tab:setting1}, we show the performance of all the methods under the evaluation of \textsc{Auc} and Accuracy when Foursquare is used as the target network and Twitter is used as the source network. By comparing method {\ourf} with method {\ourh} and comparing method {\ourtradsp} with method {\ourtrad}, we could find that sampling the old users' information could improve our prediction performance. By comparing method {\ourtrad} with {\ourh} and {\ouri}, we find that the performance of {\ourtrad} is even worse that {\ourh}, which means that old users' information without sampling could degrade the prediction performance. Comparison of {\ourtradsp} with {\ourf} and {\ouri} reveals that using the sampled old users' and the new users' information simultaneously could lead to a better prediction results. As the information owned by these new users increases, i.e., the remaining information ratio increases, the effectiveness of sampling decreases continuously as the new users and old users without sampling are becoming more and more similar. By comparing the results of methods {\ourb}, {\ourd} and {\ourtrad}, we find that using two networks at the same time could achieve better performance that using a single one. {\our} performs better that {\ourb} indicates that personalized sampling could still work when two aligned networks are used simultaneously.

From the result, we also could find that most of these methods will fail to work because of the cold start problem when remaining information ratio is 0.0, which means that the users are brand new. However, method {\our}, {\ourb}, {\ourd} and {\ournaive} could still work well because these method could get information about new users from another aligned source network. It could support the intuition of this paper that using another aligned network could help cure the cold start problem. In Table~\ref{tab:setting2}, similarly results could be gotten when Twitter is used as the target network and Foursquare is used as the aligned source network. 

\section{Related Work}
\label{sec:relatedwork}
Link prediction and recommendation first proposed in\cite{LK03} is such a significant task in graph mining that it provides researchers with the opportunities to study the network. Hasan et al. \cite{HCSZ06} is the first to study the link prediction problem as a supervised problem. Today, many social networks are heterogeneous and to conduct the link prediction in these networks, Sun et al. \cite{SBGAH11} propose a meta path-based prediction model to predict co-author relationship in the heterogeneous bibliographic network.

Mosting existing researches approach link prediction problem with a single source of information. However, Tang et al. \cite{TLK12} focus on inferring the particular type of links over multiple heterogeneous networks and develop a framework for classifying the type of social ties. To deal with the differences in information distributions of multiple networks, Qi et al. \cite{QAH13} propose to use biased cross-network sampling to do link prediction across networks. Kong et al. \cite{KZY13} propose to infer the anchor links between two heterogeneous networks with a two-phase methods.

Location-based social networks(LBSNs) are becoming quite hot in recent years and many works have been done on predicting links on these networks. Scellato et al. \cite{SNM11} predict social links by using heterogeneous information in the network. Wang et al. \cite{WPSGB11} try to predict social links by considering the moving pattern of users. 

Cold start link prediction problem start to appear in recently years. Leroy et al. \cite{LCB10} propose a two-phase method based on the bootstrap probability graph to deal with the problem with auxiliary information. Ge et al. \cite{GZ12} solve similar problem with similar methods. But they assume that most of the links are missing while multiple heterogeneous information sources are available, which doesn't conform with real cold start problem in reality and they are called pseudo cold start problem as a result.

\section{Conclusion}\label{sec:conclusion}
In this paper, we study the link prediction problem for new users and propose a supervised method {\our} to solve this problem by using information in multiple aligned heterogeneous social networks. A within-network personalized sampling method is proposed to address the differences in information distributions of new users and old users. Information from the aligned source network and that owned by the old users in the target network is transferred to help improve the prediction result.  Extensive experiments results show that {\our} works well for users of different degrees of novelty and can also solve the cold start problem.

\end{document}